\documentclass[prb,twocolumn,epsfig,epsf]{revtex4}
\usepackage{amsmath,amsthm,amsfonts,amscd,mathrsfs,pifont,bm}
\usepackage{eucal,graphicx,color}
\usepackage{esint}

\makeatletter
\@for\next:={int,iint,iiint,iiiint,dotsint,oint,oiint,sqint,sqiint,
  ointctrclockwise,ointclockwise,varointclockwise,varointctrclockwise,
  fint,varoiint,landupint,landdownint}\do{%
    \expandafter\edef\csname\next\endcsname{%
      \noexpand\DOTSI
      \expandafter\noexpand\csname\next op\endcsname
      \noexpand\ilimits@
    }%
  }
\makeatother

\newcommand*\rfrac[2]{{}^{#1}\!/_{#2}}

\begin{document}

\title  {Power Spectrum of Long Eigenlevel Sequences in Quantum Chaotic Systems}

\author {Roman Riser$^1$, Vladimir Al.~Osipov$^2$, and Eugene Kanzieper$^1$}

\affiliation
       {
       $^1$ Department of Applied Mathematics, H.I.T.--Holon Institute of
       Technology, Holon 5810201, Israel\\
       $^2$ Division of Chemical Physics, Lund University, Getingev{\"a}gen 60, Lund 22241, Sweden
       }
\date   {March 23, 2017}

\begin  {abstract}
We present a nonperturbative analysis of the power spectrum of energy level fluctuations in fully chaotic quantum structures. Focusing on systems with broken time-reversal symmetry, we employ a finite-$N$ random matrix theory to derive an exact multidimensional integral representation of the power- spectrum. The $N\rightarrow \infty$ limit of the exact solution furnishes the main result of this study -- a universal, parameter-free prediction for the power spectrum expressed in terms of a fifth Painlev\'e transcendent. Extensive numerics lends further support to our theory which, as discussed at length, invalidates a traditional assumption that the power spectrum is merely determined by the spectral form factor of a quantum system.
\end{abstract}

\maketitle
\newpage
{\it Introduction.}---Spectral fluctuations of quantum systems reflect the nature -- regular or chaotic -- of their underlying classical dynamics \cite{B-1987}. In case of fully chaotic classical dynamics, hyperbolicity (exponential sensitivity to initial conditions) and ergodicity (typical classical trajectories fill out available phase space uniformly) make quantum properties of chaotic systems universal \cite{BGS-1984}. At sufficiently long times \cite{LT} $t > T_*$, the single particle dynamics is governed by global symmetries rather than by system specialties and is accurately described by the random matrix theory \cite{M-2004,PF-book} (RMT). The emergence of universal statistical laws, anticipated by Bohigas, Giannoni and Schmit \cite{BGS-1984}, has been established within the semiclassical approach \cite{RS-2002} which links correlations in quantum spectra to correlations between periodic orbits in the associated classical geodesics. The time scale $T_*$ of compromised spectral universality is set by the period $T_1$ of the shortest closed orbit and the Heisenberg time $T_{\rm H}$, such that $T_1 \ll T_* \ll T_{\rm H}$.

Several statistical measures of level fluctuations have been devised in quantum chaology. {\it Long-range} correlations of eigenlevels on the unfolded energy scale \cite{M-2004} are measured by the variance $\Sigma^2(L)={\rm var}[{\mathcal N}(L)]$ of number of levels ${\mathcal N}(L)$ in the interval of length $L$ and by (closely related) spectral rigidity \cite{SR} $\Delta_3 (L)$. Both statistics probe the two-level correlations only and exhibit \cite{B-1985} a universal RMT behavior provided the interval $L$ is not too long, $1 \ll L \ll T_{\rm H}/T_1$. For one, the number variance is described by the logarithmic law
\begin{eqnarray} \label{nv}
    \Sigma^2_{\rm chaos}(L) = \frac{2}{\pi^2\beta} \log L + {\mathcal O}(1),
\end{eqnarray}
which indicates presence of the long-range repulsion between eigenlevels. Here, $\beta=1,2$ and $4$ denote the Dyson symmetry index \cite{M-2004}. For more distant levels, $L \gg T_{\rm H}/T_1$, system-specific features show up: both statistics display quasirandom oscillations with wavelengths being inversely proportional to periods of short closed orbits.

Individual features of quantum chaotic systems become less pronounced in spectral measures that probe the {\it short-range} fluctuations as these are largely determined by the long periodic orbits. The distribution of level spacing between (unfolded) consecutive eigenlevels, $P(s) = \langle \delta(s - E_j + E_{j+1})\rangle$, is the most commonly used short-range statistics. At small spacings, $s \ll 1$, it is mostly contributed by the {\it two-point} correlations, showing the phenomenon of symmetry-driven level repulsion, $P(s) \propto s^\beta$. (In a simple-minded fashion, this result can be read out from the Wigner surmise \cite{M-2004}). As the $s$ grows, the spacing distribution becomes increasingly influenced by spectral correlation functions of {\it all} orders. In the universal regime ($s \lesssim T_{\rm H}/T_*$), these are best accounted for by the RMT machinery which produces parameter-free (but $\beta$-dependent) representations of level spacing distributions in terms of Fredholm determinants/Pfaffians and Painlev\'e transcendents. For quantum chaotic systems with broken time-reversal symmetry $(\beta=2)$ -- that will be the focus of our study -- the level spacing distribution is given by the famous Gaudin-Mehta formula \cite{JMMS-1980,PF-book}
\begin{eqnarray}\label{LSD-PV}
    P_{\rm chaos}(s) = \frac{d^2}{ds^2} \exp \left(
            \int_{0}^{2\pi s} \frac{\sigma_0(t)}{t} dt
        \right).
\end{eqnarray}
Here, $\sigma_0(t)$ is the fifth Painlev\'e transcendent defined as the solution to the nonlinear equation $(\nu=0)$
\begin{equation} \label{PV-family}
    (t \sigma_\nu^{\prime\prime})^2 + (t\sigma_\nu^\prime -\sigma_\nu) \left(
            t\sigma_\nu^\prime -\sigma_\nu + 4 (\sigma_\nu^\prime)^2
        \right) - 4 \nu^2 (\sigma_\nu^\prime)^2 = 0
\end{equation}
subject to the boundary condition $\sigma_0(t)\sim -t/2\pi$ as $t\rightarrow~0$.

The universal RMT laws [Eqs.~(\ref{nv}) and (\ref{LSD-PV})] do not apply to quantum systems with completely integrable classical dynamics. Following Berry and Tabor \cite{BT-1977}, these belong to a different universality class -- the one shared by the Poisson point process. In particular, level spacings in a generic integrable quantum system exhibit statistics of waiting times between consecutive events in a Poisson process. This leads to the radically different fluctuation laws: the number variance $\Sigma^2_{\rm int}(L)=L$ is no longer logarithmic while the level spacing distribution $P_{\rm int}(s) = e^{-s}$ becomes exponential, with no signatures of level repulsion whatsoever. Such a selectivity of short- and long-range spectral statistical measures has long been used to uncover underlying classical dynamics of quantum systems.

{\it Power spectrum: Definition and early results.}---To obtain a more accurate characterization of the quantum chaos, it is advantageous to use spectral statistics which probe the correlations between {\it both} nearby and distant eigenlevels. Such a statistical indicator has been suggested in Ref. \cite{RGMRF-2002}. Interpreting a sequence of ordered unfolded eigenlevels $\{E_1 \le \dots \le E_M\}$ of the length $M \gg 1$ as a discrete-time random process, these authors defined its average {\it power spectrum}\cite{O-1987}
\begin{eqnarray} \label{PS-def}
    S_M(\omega) =  \frac{1}{M} \sum_{\ell=1}^M \sum_{m=1}^M \langle
                \delta E_\ell \delta E_m
        \rangle \, e^{i \omega (\ell-m)} 
\end{eqnarray}
in terms of discrete Fourier transform of 
the correlator $\langle \delta E_\ell \delta E_m \rangle$ of level displacements $\{\delta E_\ell \equiv E_\ell - \langle E_\ell\rangle\}$ on the unfolded scale ($\langle E_\ell \rangle = \ell \Delta$ with $\Delta=1$). Notice that {\it dimensionless} frequencies therein belong to the interval $0\le \omega \le \omega_{\rm Ny}$, where $\omega_{\rm Ny}\equiv\pi$ is the Nyquist frequency. A discrete nature of a so-defined random process allows us to restrict $\omega$'s to a finite set $\omega_k = 2\pi k/M$ with \cite{Even} $k \in \{1,2,\dots,M/2\}$.

Considering Eq.~(\ref{PS-def}) through the prism of a semiclassical approach, one readily realizes that, at low frequencies $\omega_k \ll T_*/T_{\rm H}$, the power spectrum is largely affected by system-specific correlations between very distant eigenlevels (accounted for by short periodic orbits). For higher frequencies, $\omega_k \gtrsim T_*/T_{\rm H}$, the contribution of longer periodic orbits becomes increasingly important and the power spectrum enters the {\it universal regime}. Eventually, in the frequency domain $T_*/T_{\rm H} \ll \omega_k \le \omega_{\rm Ny}$, long periodic orbits win over and the power spectrum gets shaped by correlations between the nearby levels.

Numerical simulations \cite{RGMRF-2002} have revealed that the average power spectrum $S_M(\omega_k)$ discriminates sharply between quantum systems with chaotic and integrable classical dynamics. While this was not completely unexpected, another finding of Ref. \cite{RGMRF-2002} came as quite a surprise:  numerical data for $S_M(\omega_k)$ could accurately be fitted by simple power-law curves, $S_M(\omega_k) \sim 1/\omega_k$ and $S_M(\omega_k) \sim 1/\omega_k^2$, for quantum systems with chaotic and integrable classical dynamics, respectively. In quantum systems with mixed classical dynamics, numerical evidence was presented \cite{GRRFSVR-2005} for the power-law of the form $S_M(\omega_k) \sim 1/\omega_k^\alpha$ with the exponent $1 <\alpha < 2$ measuring a `degree of chaoticity'. Experimentally, the $1/\omega_k$ noise was measured in Sinai microwave billiards \cite{FKMMRR-2006} and microwave networks \cite{BYBLDS-2016}. For the power spectrum analysis of Fano-Feshbach resonances in an ultracold gas of Erbium atoms \cite{FMAFBMPK-2014}, the reader is referred to Ref. \cite{PM-2015}.

For quantum chaotic systems, the universal $1/\omega_k$ law for the average power spectrum in the frequency domain $T_*/T_{\rm H}\lesssim \omega_k \ll 1$ can be read out from the existing RMT literature. Indeed, defining a set of discrete Fourier coefficients
\begin{eqnarray} \label{FK-def}
    a_k = \frac{1}{\sqrt{M}} \sum_{\ell=1}^M \delta E_\ell \, e^{i \omega_k \ell}
\end{eqnarray}
of level displacements $\{\delta E_\ell\}$, one observes the relation $S_M(\omega_k)={\rm var}[a_k]$. Statistics of the Fourier coefficients $\{a_k\}$ were studied in some detail \cite{W-1987} within the Dyson's Brownian motion model \cite{D-1962}. In particular, it is known that, in the limit $k \ll M$, they are independent Gaussian distributed random variables with zero mean and the variance ${\rm var}[a_k] = M/(2\pi^2 \beta k)$. This implies
\begin{eqnarray} \label{BM}
    S_M(\omega_k)\simeq S_\infty^{(0)}(\omega_k)= \frac{1}{\pi \beta \omega_k}
\end{eqnarray}
in concert with numerical findings. For larger $k$ (in particular, for $k \sim M$), fluctuation properties of the Fourier coefficients $\{a_k\}$ are unknown.

An attempt to determine $S_M(\omega_k)$ for higher frequencies up to $\omega_k = \omega_{\rm Ny}$ was undertaken in Ref. \cite{FGMMRR-2004} whose authors claimed to express the large-$M$ power spectrum in the entire domain $T_*/T_{\rm H}\lesssim \omega_k \le \omega_{\rm Ny}$ in terms of the spectral form factor \cite{M-2004} of a quantum system. (A similar framework was also used in subsequent papers \cite{GG-2006,RMRFM-2008}). Even though numerical simulations seemed to confirm a theoretical curve derived in Ref. \cite{FGMMRR-2004}, we believe that the status of their heuristic approach needs to be clarified.

{\it Power spectrum beyond the two-level correlations.}---In this Letter, we revisit the problem of calculating the power spectrum in quantum chaotic systems as the central assumption of previous studies \cite{FGMMRR-2004,GG-2006,RMRFM-2008} -- that at $M \gg 1$ the power spectrum is merely dominated by the two-level correlations -- cannot be substantiated. In fact, the opposite is true: as $\omega_k$ grows,
the power spectrum becomes increasingly influenced by spectral correlation functions of {\it all} orders. This is best exemplified by the large--$M$ formula
\begin{equation} \label{PS-many-point}
    S_M(\omega_k) = \frac{M}{\omega_k^2} \iint \limits_{0}^{M} dE dE^\prime e^{i\omega_k (E-E^\prime)}
    R_M(\omega_k; E, E^\prime)
\end{equation}
that follows directly from the definition Eq.~(\ref{PS-def}). Here, $R_M(\omega_k; E, E^\prime)= \langle \delta \varrho_M(\omega_k; E) \delta \varrho_M^*(\omega_k; E^\prime) \rangle$ is a one-parameter extension of the density-density correlation function $R_M(0; E,E^\prime)=\langle \delta \varrho_M(E) \delta \varrho_M(E^\prime) \rangle$ with
\begin{eqnarray}
    \delta \varrho_M(\omega_k;E) = \delta \varrho_M(E) \, \exp\Big( i \omega_k M \int_{0}^{E} d\mu \, \delta \varrho_M(\mu) \Big).
\end{eqnarray}
Contrary to $R_M(0; E, E^\prime)$, the correlator $R_M(\omega_k; E, E^\prime)$ is influenced by spectral correlation functions of all orders; so is the power spectrum.

Looking for the singular part of the power spectrum at small $\omega_k$, it is tempting to discard the higher-order correlations in Eq.~(\ref{PS-many-point}). Such a move would result in the simple approximation \cite{FGMMRR-2004} $S_M(\omega_k \ll 1) \simeq \omega_k^{-2} K_M(\omega_k/2\pi)$, where
\begin{eqnarray}
    K_M(\tau) = \frac{1}{M} \iint \limits_{0}^{M} dE dE^\prime e^{2i\pi \tau (E-E^\prime)}
    R_M(0; E, E^\prime)
\end{eqnarray}
is the spectral form factor of a quantum system. For quantum chaotic systems, known to be described by the form factor \cite{M-2004,RS-2002,B-1985} $K_M(\tau \ll 1) \simeq 2\tau/\beta$, the above approximation does reproduce the Brownian motion result Eq.~(\ref{BM}) for the power spectrum. However, for generic quantum systems with completely integrable classical dynamics, characterized by \cite{M-2004} $K_M(\tau)=1$, the very same approximation yields a wrong estimate for the power spectrum: $S_M(\omega_k \ll 1) \simeq 1/\omega_k^2$ instead of $2/\omega_k^2$. To account for the missing factor $2$, contributions of correlation functions {\it beyond} the second order should be accommodated \cite{P-sp,ROK-2017}. The latter observation not only questions the validity of the `form-factor approximation', but it also suggests that the power spectrum keeps record of spectral correlation functions of all orders.

In this Letter we show how they can be summed up {\it exactly} within the RMT approach for both finite and infinite eigenlevel sequences. In the case of broken time-reversal symmetry $(\beta=2)$, our nonperturbative theory produces a parameter-free prediction for the power spectrum in the form
\vspace{-0.5cm}
\begin{widetext}
\begin{eqnarray} \label{PS-exact}
    S_\infty(\omega_k) = {\mathcal A}(\varpi_k) \left\{{\rm Im} \int_{0}^{\infty} \frac{d\lambda}{2\pi} \, \lambda^{1-2\varpi_k^2} \, e^{i\varpi_k \lambda}
    \left[
        \exp \left(
                    - \int_{\lambda}^{\infty} \frac{dt}{t} \left( \sigma(t;\varpi_k) - i \varpi_k t + 2\varpi_k^2\right)
            \right) -1
        \right] + {\mathcal B}(\varpi_k)\right\},
\end{eqnarray}
\end{widetext}
where $\varpi_k = \omega_k/2\pi$ is a rescaled frequency, the functions ${\mathcal A}$ and ${\mathcal B}$ are defined as
\begin{subequations} \label{AB}
  \begin{align} \label{a-f}
    {\mathcal A}(\varpi_k) &= \frac{1}{2\pi} \frac{\prod_{j=1}^2 G(j+\varpi_k) G(j-\varpi_k)}{\sin(\pi \varpi_k)}, & \\
    \label{b-f}
    {\mathcal B}(\varpi_k) &= \frac{1}{2\pi} \sin(\pi \varpi_k^2)\, \varpi_k^{2\varpi_k^2-2}\, \Gamma(2-2\varpi_k^2), &
  \end{align}
\end{subequations}
$G$ is the Barnes $G$-function, $\Gamma$ is the gamma function, while $\sigma(t;\varpi_k) = \sigma_1(t)$ is a solution of the Painlev\'e~V equation Eq.~(\ref{PV-family}) at $\nu=1$
satisfying the boundary condition \cite{BC-remark,CK-2015} $\sigma_1(t) \sim i \varpi_k t -2 \varpi_k^2$ as $t\rightarrow \infty$.

The universal law Eq.~(\ref{PS-exact}), describing the power spectrum in quantum chaotic systems with broken time-reversal symmetry in the domain of its universality $T_*/T_{\rm H} \lesssim \omega_k < \omega_{\rm Ny}=\pi$, is the {\it main result} of our study. It can be regarded as a power spectrum analog of the Gaudin-Mehta formula [Eq.~(\ref{LSD-PV})].

In the low-frequency domain $\omega_k \ll 1$, the power spectrum behavior can be made explicit. Analyzing a small--$\varpi_k$ solution to Eq.~(\ref{PV-family}) at $\nu=1$, we derive:
\begin{eqnarray} \label{small-omega}
    S_\infty (\omega_k) &=& \frac{1}{4\pi^2 \varpi_k} + \frac{1}{2\pi^2} \varpi_k \log \varpi_k  \nonumber\\
         &+& \frac{\varpi_k}{12} + {\mathcal O}
    \left(
        \varpi_k^2 \log \varpi_k
        \right).
\end{eqnarray}
The leading-order term of this expansion is consistent with the $\beta=2$ Brownian-motion result Eq.~(\ref{BM}). Corrections to it have nothing in common with the results of the earlier approach \cite{FGMMRR-2004} hereby suggesting that higher-order correlations influence the power spectrum even in the low-frequency domain.

\begin{figure}[b]
\includegraphics[width=0.4\textwidth]{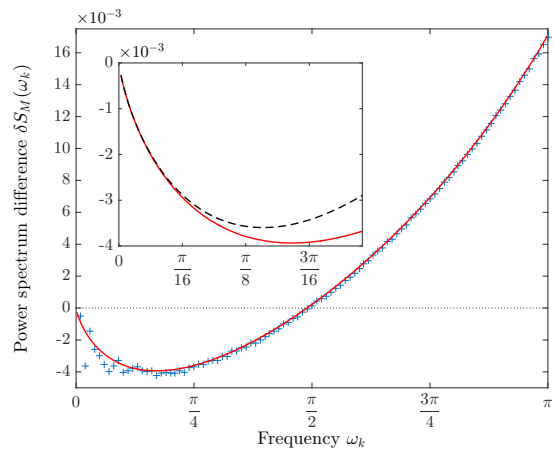}
\caption{The difference $\delta S_\infty(\omega_k) = S_\infty(\omega_k) - S_\infty^{(0)}(\omega_k)$ between the power spectrum $S_\infty(\omega_k)$ and its singular part $S_\infty^{(0)}(\omega_k) = (2\pi\omega_k)^{-1}$ as described by Eq.~(\ref{BM}) at $\beta=2$. Crosses correspond to $\delta S_M(\omega_k)$ computed for sequences of $200$ unfolded CUE eigenvalues averaged over $4 \times 10^6$ realizations. Solid line: analytical prediction for $\delta S_M(\omega_k)$ calculated through ${\rm dP}_{\rm V}$ equations with\cite{P-6-5-Remark,FM-2015} $M=1000$ [see discussion below Eq.~(\ref{pvi-bc})]. Dashed line in the inset displays the difference $\delta S_\infty(\omega_k)$ calculated using the small--$\omega_k$ expansion Eq.~(\ref{small-omega}).}\label{Fig-1}
\end{figure}

In Fig.~\ref{Fig-1}, we confront our parameter-free prediction Eq.~(\ref{PS-exact}) with the results of numerical simulations for unfolded spectra of large--$M$ circular unitary ensemble (CUE). Referring the reader to the figure caption for further details, we plainly conclude that the agreement between the theory and numerics is nearly perfect. Yet, our simulations indicate that a deviation of the simple $S_\infty^{(0)}(\omega_k)$--curve from the exact formula Eq.~(\ref{PS-exact}) does not exceed $5\%$ in the region $\omega_k < \rfrac{2}{3}\, \omega_{\rm Ny}$. For higher frequencies, the relative error quickly increases reaching its maximal value ($34\%$) at $\omega=\omega_{\rm Ny}$. This estimate should have important implications for analysis of the experimental data, see Refs. \cite{FKMMRR-2006,BYBLDS-2016,PM-2015}.

{\it Sketch of the derivation.}---To mimic an unfolded spectrum of quantum chaotic systems with broken time-reversal symmetry, we define the {\it tuned} CUE ensemble (${\rm TCUE}_{N}$) through the joint probability density function (JPDF)
\begin{equation}\label{T-CUE}
       P_N({\bm \theta}) = \frac{1}{(N+1)!}
       \prod_{\ell>m} \left|e^{i\theta_\ell} - e^{i\theta_m}\right|^2
       \prod_{\ell=1}^N \left|1-e^{i\theta_\ell}\right|^2
\end{equation}
of its $N$ eigenangles ${\bm \theta} = \{ \theta_1,\dots,\theta_N\}$. Equation (\ref{T-CUE}) can be viewed as the JPDF for traditional ${\rm CUE}_{M}$ ensemble, $M=N+1$, whose lowest eigenangle is conditioned to stay at $\theta=\theta_0=0$. Such a minor tuning of the CUE induces useful constraints \cite{ROK-2017} on the averages of and correlations between {\it ordered} ${\rm TCUE}_N$ eigenangles $\{\theta_1 \le \dots \le \theta_N\}$. Two of them, (i) $\langle \theta_\ell \rangle = \ell\,\Delta$ and (ii) $\langle (\delta\theta_\ell - \delta\theta_m)^2 \rangle =
\langle \delta\theta_{|\ell-m|}^2 \rangle$, are of particular importance. (Here, angular brackets $\langle \dots\rangle$ denote averaging with respect to the JPDF Eq.~(\ref{T-CUE}) while $\Delta = 2\pi/M$ is the mean level spacing.)

The first constraint (i) leads us to identify a set of ordered unfolded eigenlevels $\{E_0 \le E_1 \le \dots \le E_N\}$ in Eq.~(\ref{PS-def}) as $E_\ell = \theta_\ell/\Delta\; \in (0,M)$. This results in the RMT power spectrum of the form
\begin{eqnarray}\label{smwk-01}
    S_M(\omega_k) = \frac{M}{4\pi^2} \sum_{\ell=1}^{M-1} \sum_{m=1}^{M-1} \langle \delta\theta_\ell \delta \theta_m \rangle
    \, z_k^{\ell-m},
\end{eqnarray}
where $z_k = e^{i\omega_k}$. The second constraint (ii) makes it possible to reduce Eq.~(\ref{smwk-01}) down to \cite{Der-remark}
\begin{eqnarray}\label{smwk-02}
    S_M(\omega_k) = \frac{M}{4\pi^2} {\rm Re\,} \left(
              z_k \frac{\partial}{\partial z_k} - M
        \right) \, \sum_{\ell=1}^{M-1} \langle \delta \theta_\ell^2 \rangle\, z_k^\ell.
\end{eqnarray}
Expressing $\langle \theta_\ell^2\rangle$ in terms of the probability density $p_\ell(\phi)$ to observe the $\ell$'th ordered eigenvalue at $\phi$,
\begin{eqnarray}
    \langle \theta_\ell^2 \rangle &=& \int_{0}^{2\pi} \frac{d\phi}{2\pi} \,\phi^2 \, p_\ell(\phi) \nonumber\\
        &=&
     - \int_{0}^{2\pi} \frac{d\phi}{2\pi} \,\phi^2 \, \sum_{j=0}^{\ell-1} \frac{\partial E_N(j;\phi)}{\partial \phi},
\end{eqnarray}
we derive the {\it exact} representation
\begin{eqnarray} \label{central}
    S_M(\omega_k) &=& \frac{M}{\pi} {\rm Re} \Bigg[
        \left(
              z_k \frac{\partial}{\partial z_k} - M
        \right)  \nonumber\\
        &\times& \frac{z_k}{1-z_k} \int_{0}^{2\pi} \frac{d\phi}{2\pi}\,\phi\, \Phi_N (\phi; 1-z_k)
    \Bigg].
\end{eqnarray}
Here,
\begin{subequations} \label{phi-n-gf}
  \begin{align} \label{ph-n-02-a}
    \Phi_N(\phi;\zeta) &= \sum_{\ell=0}^N (1-\zeta)^\ell E_N(\ell; \phi) & \\
    \label{phi-n-02}
     &= \left(
   \int_0^{2\pi} - \zeta \int_{0}^{\phi}  \right) \prod_{\ell=1}^N \frac{d\theta_\ell}{2\pi}
   P_N({\bm \theta}) &
  \end{align}
\end{subequations}
is the generating function \cite{PF-book} of the probabilities $E_N(\ell;\phi)$ to find exactly $\ell$ eigen-angles in the interval $(0,\phi)$. Equations (\ref{central}), (\ref{phi-n-gf}) and (\ref{T-CUE}) are central to our nonperturbative theory of the power spectrum.

{\it Painlev\'e VI representation.}---Multidimensional integrals of the CUE type [Eqs.~(\ref{phi-n-02}) and (\ref{T-CUE})] have been studied in much detail in Ref. \cite{FW-2004}. Building upon it, we can represent the finite-$M$ power spectrum Eq.~(\ref{central}) in terms of the sixth Painlev\'e function $\tilde{\sigma}_N$ such that
\begin{equation} \label{phin}
    \Phi_N(\phi;\zeta) = \exp \left(
            -\int_{\cot(\phi/2)}^{\infty} \frac{dt}{1+t^2} \left( \tilde{\sigma}_N(t;\zeta) + t \right)
        \right).
\end{equation}
The function $\tilde{\sigma}_N(t;\zeta)$ is a solution to the Painlev\'e VI equation
\begin{eqnarray} \label{pvi}
    \left( (1+t^2)\,\tilde{\sigma}_N^{\prime\prime} \right)^2 &+& 4 \tilde{\sigma}_N^\prime (\tilde{\sigma}_N - t \tilde{\sigma}_N^\prime)^2 \nonumber\\
    &+& 4 (\tilde{\sigma}_N^\prime+1)^2 \left(
        \tilde{\sigma}_N^\prime + (N+1)^2
    \right) = 0\qquad
\end{eqnarray}
subject to the boundary condition \cite{ROK-2017}
\begin{eqnarray} \label{pvi-bc}
    \tilde{\sigma}_N(t;\zeta) = -t + \frac{N(N+1)(N+2)}{3\pi t^2} \zeta + {\mathcal O}(t^{-4})
\end{eqnarray}
as $t\rightarrow \infty$. Equations (\ref{central}) and (\ref{phin})--(\ref{pvi-bc}) provide an exact RMT solution for the power spectrum at finite $N$. Alternatively, but equivalently, the exact solution for $S_M(\omega_k)$ can be formulated \cite{ROK-2017} in terms of discrete Painlev\'e~V equations (${\rm dP}_{\rm V}$), much in line with Ref.~\cite{FW-2003-05}. The latter representation is particularly useful for efficient numerical evaluation of the power spectrum for relatively large values of $N$; such a computation, displayed in Fig. \ref{Fig-1}, shows a remarkable agreement of our exact solution with the power spectrum determined for numerically {\it simulated} CUE spectra.

{\it Asymptotic analysis of the exact solution.}---To identify a universal, parameter-free, law for the power spectrum, expected to emerge in the limit $N \rightarrow \infty$, an asymptotic analysis of the exact solution Eqs.~(\ref{central}) and (\ref{phin})--(\ref{pvi-bc}) is required. To facilitate such an analysis, we notice that the generating function $\Phi_N(\phi;\zeta)$ [Eq.~(\ref{phi-n-02})] entering the exact solution Eq.~(\ref{central}) with $\zeta = 1 - z_k$ admits a representation
\begin{eqnarray} \label{T-det-01}
     \Phi_N(\phi; 1 - z_k) = \frac{e^{i\phi \tilde{\omega}_k N}}{N+1} \, D_N[f_{\tilde{\omega}_k}(z;\phi)]
\end{eqnarray}
in terms of the Toeplitz determinant
\begin{equation}\label{T-det-02}
    D_N[f_{\tilde{\omega}_k}(z;\phi)] =  {\rm det} \left( \frac{1}{2 i\pi} \oint_{|z|=1} \frac{dz}{z} \,z^{\ell-j}
    f_{\tilde{\omega}_k}(z;\phi) \right)
\end{equation}
$(0 \le j,\ell \le N-1)$ whose Fisher-Hartwig symbol
\begin{eqnarray} \label{FHS}
    f_{\tilde{\omega}_k}(z;\phi) = |z-z_1|^2 \left(\frac{z_2}{z_1}\right)^{\tilde{\omega}_k} g_{z_1,\tilde{\omega}_k} (z)\, g_{z_2,-\tilde{\omega}_k}(z)
\end{eqnarray}
possesses a power-type singularity at $z=z_1 = e^{i\phi/2}$ and two jump discontinuities
\begin{eqnarray}
    g_{z_{j}, \pm{\tilde{\omega}_k}} (z) = \left\{
                                         \begin{array}{ll}
                                           e^{\pm i\pi \tilde{\omega}_k}, & \hbox{$0 \le {\rm arg\,} z < {\rm arg\,} z_{j}$} \\
                                           e^{\mp i\pi \tilde{\omega}_k}, & \hbox{${\rm arg\,} z_{j} \le {\rm arg\,} z < 2\pi$}
                                         \end{array}
                                       \right.
\end{eqnarray}
at $z = z_{1,2}$ with $z_2 = e^{i(2\pi -\phi/2)}$. (Here, we have followed the standard terminology and notation, see e.g. Ref. \cite{DIK-2014}).

To perform the integral in Eq.~(\ref{central}) in the large-$N$ limit, one needs to know uniform asymptotics of the Toeplitz determinant Eq.~(\ref{T-det-02}) in the subtle case of merging singularities. These particular asymptotics of Toeplitz determinants were recently studied in great detail by Claeys and Krasovsky \cite{CK-2015}. Making use of their Theorems 1.5 and 1.11, we end up -- after a somewhat involved calculation \cite{ROK-2017} -- with the universal law for the power spectrum announced in Eqs.~(\ref{PS-exact}) and (\ref{AB}) above.

{\it Summary.}---We revisited the problem of calculating the power spectrum in quantum chaotic systems putting particular emphasis on the shortcomings of widely used `form factor' approximation. Having argued that the power spectrum is shaped by spectral correlations of all orders, we showed how their contributions can be summed up nonperturbatively within the RMT approach. Our main result -- a parameter-free prediction for the power spectrum expressed in terms of a fifth Painlev\'e transcendent [Eq.~(\ref{PS-exact})] -- is expected to hold universally, at not too low frequencies, for a variety of quantum systems with chaotic classical dynamics and broken time-reversal symmetry. Meticulous numerical analysis revealed that the heuristically anticipated \cite{RGMRF-2002,FGMMRR-2004} $1/\omega_k$ law for the power spectrum can deviate quite significantly from our exact solution. This will have important implications for analysis of the experimental data.

This work was supported by the Israel Science Foundation through the Grant No. 647/12. V.~Al.~O. acknowledges support from NanoLund and the Knut and Alice Wallenberg Foundation (KAW).


\vspace{-0.5cm}

\begin{references}

\bibitem{B-1987} M. V. Berry, Proc. R. Soc. A {\bf 413}, 183 (1987).

\bibitem{BGS-1984} O.~Bohigas, M.~J.~Giannoni, and C.~Schmit, Phys.
    Rev. Lett. {\bf 52}, 1 (1984).

\bibitem{LT} Long times correspond to short correlation scales on the energy axis.

\bibitem{M-2004} M. L. Mehta, {\it Random Matrices} (Elsevier, Amsterdam, 2004).

\bibitem{PF-book} P.~J.~Forrester, {\it Log-Gases and Random Matrices} (Princeton University Press, Princeton NJ, 2010).

\bibitem{RS-2002} K.~Richter and M.~Sieber,
    Phys. Rev. Lett. {\bf 89}, 206801 (2002);
    S.~M\"uller, S.~Heusler, P.~Braun, F.~Haake, and A.~Altland, Phys. Rev. Lett. {\bf 93}, 014103 (2004);
    S.~Heusler, S.~M\"uller, A.~Altland, P.~Braun, and F.~Haake,
    Phys. Rev. Lett. {\bf 98}, 044103 (2007); S.~M\"uller,
    S.~Heusler, A.~Altland, P.~Braun, and F.~Haake, New J. Phys. {\bf 11}, 103025
    (2009).

\bibitem{SR} Spectral rigidity $\Delta_3(L)$ is the least-square deviation of the spectral counting function ${\mathcal N}(E)$ from the best fit to a straight line over the interval of the length $L$. See Ref. \cite{M-2004}.

\bibitem{B-1985} M.~V.~Berry, Proc. R. Soc. A {\bf 400}, 229 (1985); Nonlinearity {\bf 1}, 399 (1988).

\bibitem{JMMS-1980} M.~Jimbo, T.~Miwa, Y.~M\^{o}ri, and M.~Sato, Physica D {\bf 1}, 80 (1980).

\bibitem{BT-1977} M.~V.~Berry and M.~Tabor, Proc. R. Soc. A {\bf 356}, 375 (1977).

\bibitem{RGMRF-2002} A.~Rela\~{n}o, J.~M.~G.~G\'{o}mez, R.~A.~Molina, J.~Retamosa, and E.~Faleiro, Phys. Rev. Lett. {\bf 89},
    244102 (2002).

\bibitem{O-1987} Similar statistics has previously been used by Odlyzko who analyzed power spectrum of the {\it spacings} between zeros of the Riemann zeta function. See: A.~M.~Odlyzko, Math. Comput. {\bf 48}, 273 (1987).

\bibitem{Even} From now on, $M$ is assumed to be an even integer.

\bibitem{GRRFSVR-2005} J.~M.~G.~G\'{o}mez, A.~Rela\~{n}o, J.~Retamosa, E.~Faleiro, L.~Salasnich, M.~Vrani\v{c}ar, and M.~Robnik, Phys. Rev. Lett. {\bf 94}, 084101 (2005); A.~Rela\~{n}o,  Phys. Rev. Lett. {\bf 100}, 224101 (2008).

\bibitem{FKMMRR-2006} E.~Faleiro, U.~Kuhl, R.~A.~Molina, L.~Mu\~{n}oz, A.~Rela\~{n}o, and J.~Retamosa, Phys. Lett. A {\bf 358}, 251 (2006).

\bibitem{BYBLDS-2016} M.~Bia\l{}ous, V.~Yunko, S.~Bauch, M.~\L{}awniczak, B.~Dietz, and L.~Sirko, Phys. Rev. Lett. {\bf 117}, 144101 (2016).

\bibitem{FMAFBMPK-2014} A.~Frisch, M.~Mark, K.~Aikawa, F.~Ferlaino, J.~L.~Bohn, C.~Makrides, A.~Petrov, and S.~Kotochigova,
    Nature {\bf 507}, 475 (2014).

\bibitem{PM-2015} J.~Mur-Petit and R.~A.~Molina, Phys. Rev. E {\bf 92}, 042906 (2015).

\bibitem{W-1987} M.~Wilkinson, J. Phys. A {\bf 21}, 1173 (1988); B.~Mehlig and M.~Wilkinson, Phys. Rev. E {\bf 63}, 045203 (2001).

\bibitem{D-1962} F.~J.~Dyson, J. Math. Phys. {\bf 3}, 1191 (1962).

\bibitem{FGMMRR-2004} E.~Faleiro, J.~M.~G.~G\'{o}mez, R.~A.~Molina, L.~Mu\~{n}oz, A.~Rela\~{n}o, and J.~Retamosa, Phys. Rev. Lett. {\bf 93},
    244101 (2004).

\bibitem{GG-2006} A.~M.~Garc\'{i}a-Garc\'{i}a, Phys. Rev. E {\bf 73}, 026213 (2006).

\bibitem{RMRFM-2008} A.~Rela\~{n}o, L.~Mu\~{n}oz, J.~Retamosa, E.~Faleiro, and R.~A.~Molina, Phys. Rev. E {\bf 77}, 031103 (2008).

\bibitem{P-sp} In case of completely integrable classical dynamics, one observes that $\langle\delta E_\ell \delta E_m\rangle = \min(\ell, m)$. Equation (\ref{PS-def}) then yields the exact result $S_M(\omega_k) = 1/2\sin^2(\omega_k/2)$. In the domain $\omega_k \ll 1$, this reduces to $S_M = 2/\omega_k^2$.

\bibitem{ROK-2017} R.~Riser, V.~Al.~Osipov, and E.~Kanzieper (unpublished).

\bibitem{BC-remark} Strictly speaking, this boundary condition is only valid for $0\le \omega_k \le \omega_{\rm Ny}/2$; for a detailed discussion the reader is referred to Ref. \cite{CK-2015}.

\bibitem{CK-2015} T.~Claeys and I.~Krasovsky, Duke Math. J. {\bf 164}, 2897 (2015).

\bibitem{P-6-5-Remark} A recent study \cite{FM-2015} suggests that replacement of $S_M(\omega_k)$ with $S_\infty(\omega_k)$ brings a relative error of the order $\mathcal{O}(M^{-2})$.

\bibitem{FM-2015} P.~J.~Forrester and A.~Mays, Proc. R. Soc. A {\bf 471}, 20150436 (2015).

\bibitem{Der-remark} Here and in Eq.~(\ref{central}) the derivative with respect to $z_k$ should be taken as if $z_k$ were a continuous variable.

\bibitem{FW-2004} P.~J.~Forrester and N.~S.~Witte, Nagoya Math.~J. {\bf 174}, 29 (2004).

\bibitem{FW-2003-05} P.~J.~Forrester and N.~S.~Witte, Nonlinearity {\bf 16}, 1919 (2003); Nonlinearity {\bf 18}, 2061 (2005).

\bibitem{DIK-2014} P.~Deift, A.~Its, and I.~Krasovsky, in: {\it Random
Matrix Theory, Interacting Particle Systems, and Integrable Systems} (Cambridge
University Press, New York, 2014), p. 93.

\end{references}
\end{document}